# Explainable Graph-based Search for Lessons-Learned Documents in the Semiconductor Industry

Hasan Abu-Rasheed[1], Christian Weber[1], Johannes Zenkert[1], Roland Krumm[2], Madjid Fathi[1]

[1] University of Siegen, Siegen, Germany
`hasan.abu.rasheed@uni-siegen.de`
[2] Elmos Semiconductor AG, Dortmund, Germany

**Abstract.** Industrial processes produce a considerable volume of data and thus information. Whether it is structured sensory data or semi- to unstructured textual data, the knowledge that can be derived from it is critical to the sustainable development of the industrial process. A key challenge of this sustainability is the intelligent management of the generated data, as well as the knowledge extracted from it, in order to utilize this knowledge for improving future procedures. This challenge is a result of the tailored documentation methods and domain-specific requirements, which include the need for quick visibility of the documented knowledge. In this paper, we utilize the expert knowledge documented in chip-design failure reports in supporting user access to information that is relevant to a current chip design. Unstructured, free, textual data in previous failure documentations provides a valuable source of lessons-learned, which expert design-engineers have experienced, solved and documented. To achieve a sustainable utilization of knowledge within the company, not only the inherent knowledge has to be mined from unstructured textual data, but also the relations between the lessons-learned, uncovering potentially unknown links. In this research, a knowledge graph is constructed, in order to represent and use the interconnections between reported design failures. A search engine is developed and applied onto the graph to answer queries. In contrast to mere keyword-based searching, the searchability of the knowledge graph offers enhanced search results beyond direct matches and acts as a mean for generating explainable results and result recommendations. Results are provided to the design engineer through an interactive search interface, in which, the feedback from the user is used to further optimize relations for future iterations of the knowledge graph.

**Keywords** Knowledge graphs, Graph-based search, Text mining, Explainable search results, Lessons learned.

## 1 Introduction and Research Purpose

Sufficient utilization of well-documented lessons-learned for an application in the domain of semiconductor design is a non-trivial task. With a multitude of design cases, their relevant failures and corresponding solutions, user's access to relevant past experiences presents a challenge in the industrial environment [1]. Traditional information retrieval, e.g. keyword-based search, is not sufficient to discover and define similarities in semi- and unstructured textual data in chip-design documentations [2]. Moreover, processing free textual input for search queries itself reveals multiple challenges, such as the multilingual nature of documentation, the domain-specific terminology, abbreviations, misspellings, grammatical mistakes and incomplete sentences.

Defining the relevancy of a current design case to other failures, which occurred in the past, requires a semantic understanding of the domain-specific description of those failures. Natural language processing (NLP) and text mining pipelines offer the first step in this understanding. However, their strength in handling textual inputs is still limited when multiple representations of information are extracted from different data sources. This can happen when integrating basic textual documentation with structural metadata representations of the chip modules and their corresponding elements. To address this challenge, a semantic definition of interlinked networks of available knowledge is needed to represent the relations between entities, which are extracted from multisource, multitype data [3], [4]. This fusion and semantic definition of knowledge entities, extracted from the lessons-learned information, is accomplished in this research through the construction of a multidimensional knowledge graph.

In the graph, not only failure cases are represented in form of lessons-learned, but also their structural information and semantic relations. A graph search functionality is then designed to boost the exploitation of previous lessons-learned for search queries and thus establish an easy access to their documentation. This is then used to provide previous expert-knowledge to new design engineers. Such search functionality is offered through an interface that supports a feedback loop for assessing retrieved results and enabling a sustainable learning and knowledge update. However, providing the feedback is also strongly influenced by user's understanding of the reasons behind retrieving a certain search result from the knowledge graph. Therefore, this research designs a graph-based explainability approach, to offer the user explanations on the reasons that led the system to consider a certain search result relevant to their search query and design scenario. Explanations are generated from the NLP functionality and the knowledge graph structure. They are provided to the users in textual and visual formats through the interface.

In the following sections of this paper, we highlight the literature background of the conducted research in section 2. In section 3, the developed methods, including system and graph structures, graph-based explainability and the interactive interface, are presented and discussed. Section 4 provides an overview on the archived results; and the work is concluded with future highlights in section 5.

## 2    Background and Related Work

Methods that are intended to analyze structured, semi-structured and unstructured textual data utilize text mining pipelines and techniques [5], [6]. Hybridizing multiple techniques is also a method that can be used for further enhancing models in the industrial context [7]. A suitable preparation of the textual data is essential to achieve the goal of text mining. Each domain-specific corpus of text requires special data pre-processing, based on the methods of documentation, data storage and the domain terminology itself [2]. Therefore, in the context of semantic representation, the utilization of domain-specific features plays a considerable role in relation extraction, and thus in constructing knowledge graphs from the information extracted from the text. An example is the work of Subasic et al. [8], where the solution to adapt the knowledge graph to the domain terminology was to build the graph on two levels: a static level, corresponding to predefined databases, and a dynamic level, that corresponds to the special terminology.

Knowledge graphs have emerged as a graphical representation of a knowledge base; and got integrated with multiple technologies and applications [9]. Among the tasks that knowledge graphs have been involved in, enhancing search and matching is a promising field, especially with the advances in search engines and semantic web. Supported by text mining, Tiwari et al. [10] developed a concept to integrate the Resource Description Framework (RDF) with natural language understanding, in order to find a common format that supports human and machine understanding together. This, in turn, supports the representation and retrieval of information from heterogeneous and distributed data sets. Our approach intersects with the previous examples in terms of utilizing domain-specific text mining in constructing the knowledge graph for search and information retrieval. However, we propose an approach to integrate domain information in the graph structure itself, in order to enable:

1) Domain-specific relation extraction alongside the text-mining-based relation extraction.
2) Domain-specific search and information retrieval.
3) An easy integration of new knowledge in the graph, supporting a dynamic process of transferring lessons-learned from the past and present to future design scenarios.
4) An explainable information retrieval process, both verbally and visually, with the utilization of NLP and the knowledge graph structure.

In order to achieve those contributions, a range of knowledge extraction, interlinking and management methods have been used, which are discussed in detail in the following section.

## 3    Search Methods

### 3.1    Multidimensional Domain-specific Knowledge Graph

In the handled industrial scenario, multiple information sources provide insights into the lessons-learned from previous design failures. In order to harvest the information from those sources and extract the knowledge within them, we design a knowledge graph structure that is capable of reflecting domain-specific features in heterogeneous data sources. Proposed knowledge graph consists of multiple types of nodes, each corresponding to a different data source, and thus, a different dimension of the available information in design processes. Those multi-type nodes are connected through text-mining-based semantic relations, in order to model the relationships between them, and enhance the visibility of the lessons-learned documents.

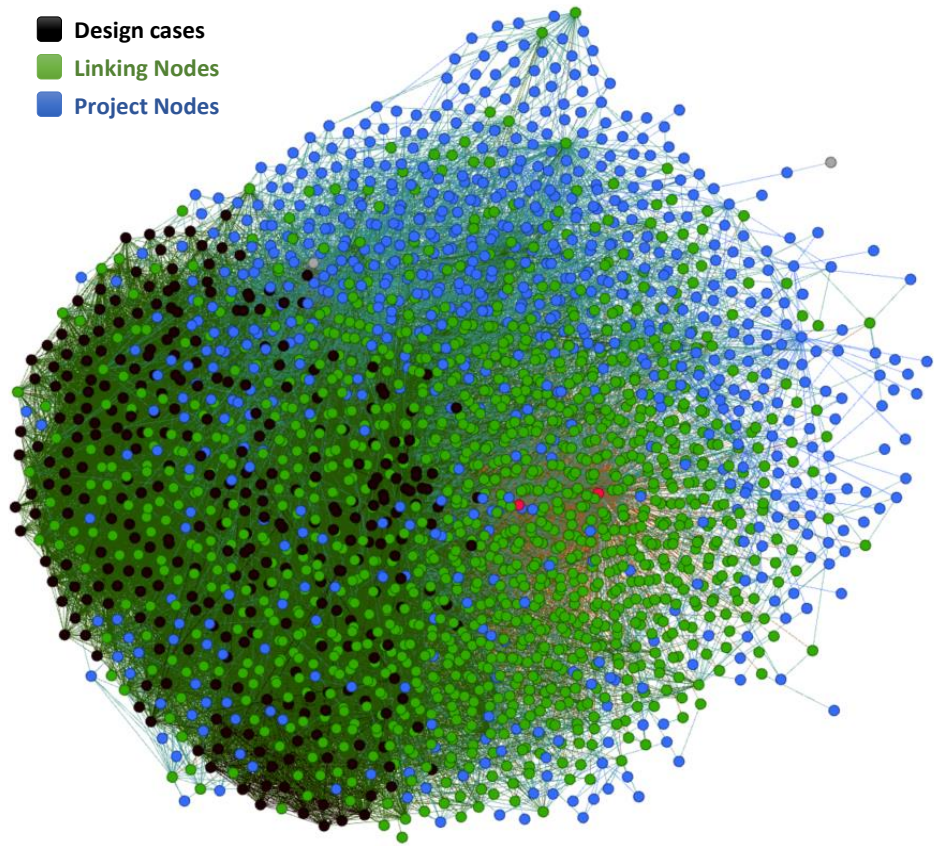

**Fig. 1** Proposed multi-node-type knowledge graph

The nodes in the proposed graph concept, see Fig.1, include:

- **Design Case Nodes:** This type of nodes corresponds to the lessons that are learned from previous design failures and solutions. The node includes detailed information about the failure description, its reasons and solution. It represents the design document that has been generated when handling the failure.
- **Project Element Nodes:** This type of nodes corresponds to the meta-data that expresses a structural dependency of design projects. It reflects the relationships between project modules and the chip elements. Relations between this type of nodes can uncover the relevance between failure cases that occur in multiple modules, whether within the same project or within multiple projects.
- **Linking Nodes:** This type of nodes is the part of the proposed solution, which tailors the graph structure itself to reflect data mining results and domain-specific features. Linking nodes play two main roles in the graph: 1) representing the text mining techniques that discover similarities between past and future design cases. 2) representing the domain-specific terminology used in the context of chip-design documentation. In this research, we use Term Frequency and Inverse Document Frequency (TFIDF) to provide a simple mean of ranking the similarities and reflecting an order of relevance [11]. To overcome the limitations of TFIDF, we utilize an N-gram approach and a specialized technical-term classifier, in order to correspond to documentation patterns and domain-specific terminology, such as the writing styles of design engineers, technical vocabulary and abbreviations. An entity recognition approach has been utilized to identify certain categories of vocabulary that are commonly used to describe failures and their solutions. A pre-study revealed a high tendency of design engineers to use certain entities as domain-dependent vocabulary in their documentations. This vocabulary emerges from the specific nature of the technical domain. Therefore, we exploit those entities in the form of highly weighted linking nodes that reflect a relational priority in the search functions within the graph. This, in turn, integrates the domain-specific information effectively in the information retrieval task.

In addition to defining graph node types, relation extraction is the following essential task that allows graph structure to represent the application domain, and thus enhance the domain-specific information retrieval. Relation extraction in the proposed approach is based on three levels:
1) Extracting relations between the lessons-leaned documents themselves.
2) Extracting relations to linking nodes and structural nodes in the graph.
3) Adding new relations that reflect newly added lessons-learned.

The first type of relations reflects the semantic inter-relations between design-failure descriptions. It inherits a cause and effect logic from the structure of the documented design reports. Relations of this type are weighted based on the TFIDF scores and the expert knowledge. The second type of relations represents the influence of domain specific terminology and project structure. It relates lessons-learned nodes to each other through the domain-specific linking nodes and project structural nodes. This type of relations has a high priority in the search and retrieval task, since domain experts have expressed a particular interest for retrieving results based on domain-specific terminology. The third type of relations represents the dynamic potential of updating the knowledge graph with accumulated design-experience through new lessons-learned. This plays the important role of ensuring a sustainable knowledge update in the organization.

Those multiple levels of relations are also designed to: 1) provide a mean of ordering retrieved search results, 2) provide multiple levels of result explanation to the user. Explainable results are meant to support the user's ability of assessing the relevancy of retrieved documents to their current design scenario. In the proposed approach, we utilize this feature to support a feedback loop, in which the expert user can effectively evaluate the retrieved result, in order to enable future relation analysis and update of the graph structure. These are continuous feedback and enhancement processes during the deployment of the system, especially with new lessons-learned being added dynamically to the graph.

## 3.2 Explainable Graph Search Functionality

The goal of the proposed solution is to support design engineers with quick accessibility to previous chip design cases, including their descriptions and solutions. Therefore, the use-case of the proposed approach takes the form of a search engine, which provides the possibility to retrieve relevant design cases to a current one. The engine allows users to search previous lessons learned with a short description of either the failure itself, the module in which the failure occurs or other failure-related details. Search functionality that is designed for this solution follows the prioritized relations in the graph, to produce an ordered list of relevant, previous, design cases. From the information included in graph's relations, search functions can retrieve results and also generate a high-level explanation regarding the reasons behind retrieving these results.

Graph-based search plays a considerable role in the proposed system for retrieving not only directly related search results, but also relevant results that are not shown to the user with traditional key-word searches. In the proposed search engine, search strategies define direct-hit results, which are retrieved from graph nodes corresponding to a direct match to the search query. The strategy then utilizes graph shortest-path calculation, starting from the direct-hit node, in order to find relevant nodes to the search query. Shortest paths represent a certain length of defined relations between two nodes. Therefore, the well-defined types of graph relations in section 3 perform a key role in finding those relevant results in the constructed graph. Both direct hits and relevant results are then ordered, equipped with a generated textual explanation, and provided to the user through the search interface, as shown in Fig.2. This allows the expert to navigate to the retrieved result, get the information they are looking for and provide any necessary feedback, if needed.

From a diverse spectrum of explanation methods, verbal and visual explanations were selected for this research. Verbal explanations were a direct result of the text analysis in the developed pipeline. It enabled the algorithm of following the feature similarity between two documents. Moreover, the structure of the knowledge graph revealed transitive relations between documents that are relevant to each other due to a third document that intermediates the two. Graph relations are also expressed verbally, alongside the document features in a template-based, human-understandable explanation. This explanation method was also supported by a visual component, which is based on the graph structure itself. It was provided to the user as a partial graph, containing their search input and all retrieved results, as an interlinked network of documents. This way, the user is able to see the relations directly, read their reasons in the graph, and easily differentiate the search direct results from the transitively retrieved ones.

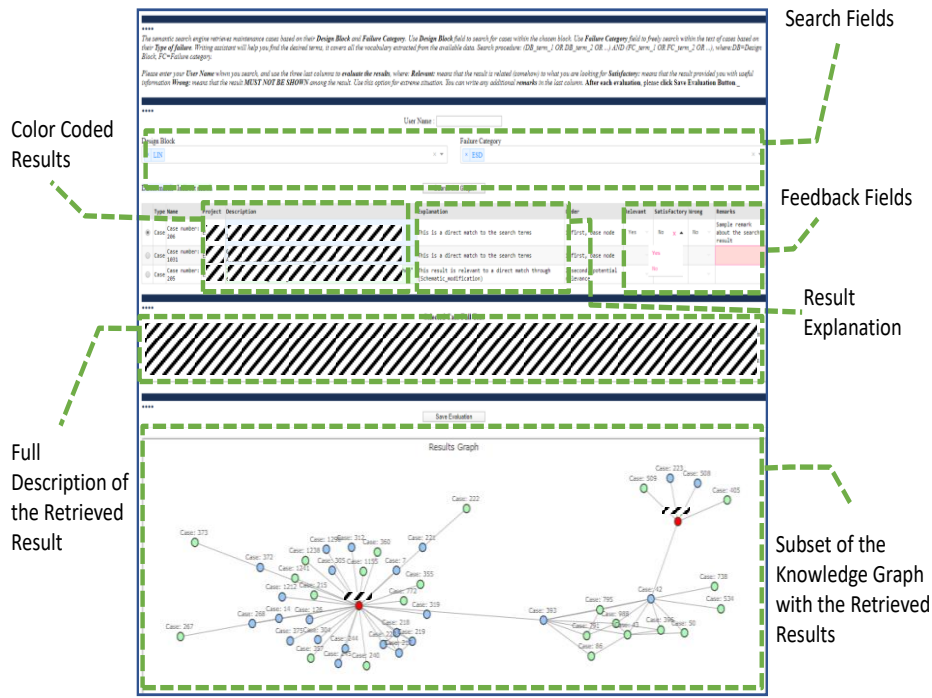

**Fig. 2** Proposed graph-based search interface

### 3.3 Search Interface and Feedback Collection

Enhancing the visibility of previous lessons-learned depends on the underlying representation methods of those lessons. In the application field of this research, expert design-engineers can utilize a filtered, use-case-driven document accessibility, which enables a goal-oriented retrieval of lessons learned. This is due to the considerable amounts of available information in the chip design scenario. Following that principle, an interactive search interface has been proposed to design engineers, supported by tailored features, such as semi-assisted, dictionary-based writing feature, color-coded results, result explanations and quick feedback fields for a continuous online evaluation. Semi-assisted search fields provide expert users with an intermediate validation of the availability of typed-in search terms. This dictionary-based method is quicker than graph scan. It evaluates the searched terms based on their potential existence in the text corpus of the lessons-learned. When a search term is missing from the corpus, the writing assistant suggests similar vocabulary that is assured to retrieve results once searched within the knowledge graph. Explanations in the interface accompany each retrieved result, in order to enable the user to better assess the suitability and relevance of the that result to their current design. Feedback fields are designed to be simple and quick, including the user's assessment of the result's relevance and value-added. This feedback can be potentially utilized in enhancing the graph structure and relations for better domain-specific retrieval.

## 4 Evaluation and Results

In order to evaluate the value-added to the design engineer, we define a set of domain-specific key performance indicators (KPIs), which reflect the need for higher visibility of information regarding similar previous design cases. Defined KPIs include quantitative measures, which reflect the change in number of results through the utilization of graph relations; as well as qualitative measures that have been defined to reflect the satisfaction level of users and the value-added through the explanations.

The experiment setup for testing the proposed system included experts in the domain and developers. Domain experts defined multiple combinations of challenging search terms, which are usually documented in different ways depending on the engineer who is writing the report. Developers have generated variations of the search terms to cover the multiple searching methods the interface offers. Search test-combinations were then fed to the system. Retrieved results were validated in terms of the defined KPIs, i.e. their numbers in comparison to key-word search (see Fig. 3), the result's relevancy to the current design scenario, the ability of the system to find semantically similar documents, as well as the reasoning and ratio of the graph-based transitively retrieved results to the direct-hits ones. Moreover, the feedback loop was evaluated by the experts on two levels: 1) the potential to assess retrieved result based on the generated explanations, and 2) the ability to dynamically modify the graph structure based on the provided feedback. The latter point was evaluated by the time needed to get an updated version of the graph, which reflects the provided feedback.

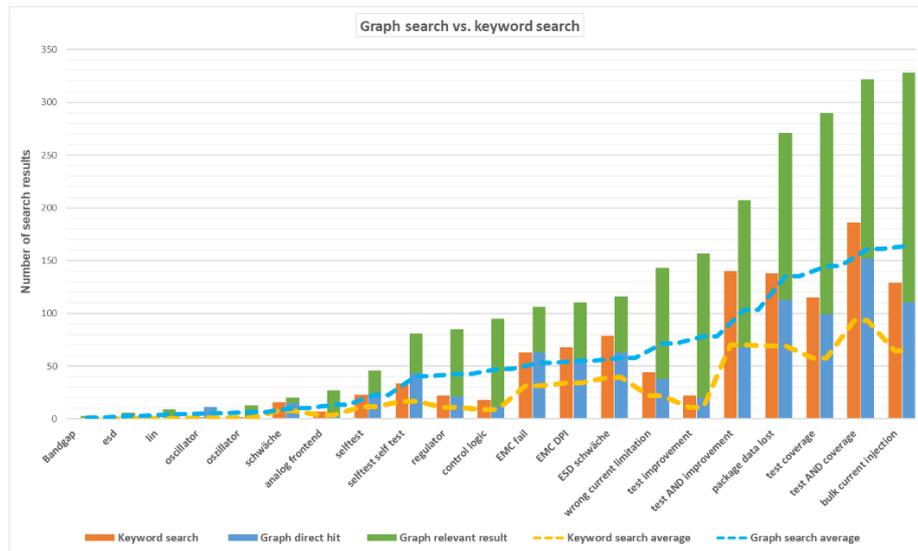

**Fig. 3** Quantitative Comparison of Search Results

Preliminary qualitative results showed that the expert users were able to reach relevant documents that are semantically related to their current design, despite the fact that failure descriptions used different vocabulary. The relation between those failures was discovered from the contextual information of the failure, which is successfully embedded in the knowledge graph. Qualitative evaluation of the system also revealed that the explanations have provided the experts with new insights of the documentation process itself. Experts reported that they were able to identify multiple documentation practices that should be enhanced in the company, in order to enhance the information retrieval in the future. An example of such practices is the use of similar, yet different, expressions to describe the failure's location on the chip.

Fig.3 represents a preliminary outlook on the quantitative change in the number of retrieved results, through the comparison between a normal key-word search in the database, and the search through extracted relations in the knowledge graph. In this figure, a group of 22 sample searches is used. The numbers show an increase of relevant results, which are retrieved from the graph-database through their semantic relations. It is noticeable also that direct search results themselves are different from the key-word results, since they are improved by the implemented text-mining techniques. On average, a 75.6% increase of result retrieval, and thus visibility to the user, is achieved with the adoption of the proposed solution.

## 5 Conclusion and Future Perspectives

In this research, a domain-specific search engine has been proposed in the field of semiconductor chip design. Proposed system is developed to support design engineers with a quick, filtered and use-case-oriented visibility of lessons-learned from previous design failures. A knowledge graph forms the base of the proposed search engine, which captures the semantic relations between information extracted from multiple documentation sources. Textual semi- and un-structured data, along with project metadata are mined and fused in one knowledge graph. The graph is then equipped with an explainable search functionality, which allows the retrieval of direct and relevant search results based on the graph semantic relations. An interactive search interface has been designed to provide color-coded search results along with their explanations, allowing users to provide their feedback on the retrieved results. Feedback loop is meant to support future enhancements of the graph relation extraction and prediction. The developed system is being tested on-site with end users to collect their quantitative and qualitative feedback.

Future steps of the development include the design of rating functions that provide more refined weights of the search results, keeping the domain-specific needs as the main focus of the information retrieval process.

**ACKNOWLEDGMENTS**
This work was supported by the EU project iDev40. This project has received funding from the ECSEL Joint Undertaking (JU) under grant agreement No 783163. The JU receives support from the European Union's Horizon 2020 research and innovation program and Austria, Germany, Belgium, Italy, Spain, Romania. The information

**References**

1. C.-F. Chien and L.-F. Chen, "Data mining to improve personnel selection and enhance human capital: A case study in high-technology industry," *Expert Syst. Appl.*, vol. 34, no. 1, pp. 280–290, Jan. 2008, doi: 10.1016/j.eswa.2006.09.003.
2. Abu-Rasheed H., Weber C., Zenkert J., Czerner P., Krumm R., Fathi M. (2021) A Text Extraction-Based Smart Knowledge Graph Composition for Integrating Lessons Learned During the Microchip Design. In: Arai K., Kapoor S., Bhatia R. (eds) Intelligent Systems and Applications. IntelliSys 2020. Advances in Intelligent Systems and Computing, vol 1251. Springer, Cham. https://doi.org/10.1007/978-3-030-55187-2_43.
3. S. Thoma, A. Rettinger, and F. Both, "Knowledge Fusion via Embeddings from Text, Knowledge Graphs, and Images," *ArXiv170406084 Cs Stat*, Apr. 2017, Accessed: Mar. 12, 2019. [Online]. Available: http://arxiv.org/abs/1704.06084.
4. L. Guo, W. Zuo, T. Peng, and L. Yue, "Text Matching and Categorization: Mining Implicit Semantic Knowledge from Tree-Shape Structures," *Math. Probl. Eng.*, vol. 2015, pp. 1–9, 2015, doi: 10.1155/2015/723469.
5. F. Shi, L. Chen, J. Han, and P. Childs, "A Data-Driven Text Mining and Semantic Network Analysis for Design Information Retrieval," *J. Mech. Des.*, vol. 139, no. 11, p. 111402, Nov. 2017, doi: 10.1115/1.4037649.
6. M. Allahyari *et al.*, "A Brief Survey of Text Mining: Classification, Clustering and Extraction Techniques," *ArXiv170702919 Cs*, Jul. 2017, Accessed: Nov. 22, 2018. [Online]. Available: http://arxiv.org/abs/1707.02919.
7. J. Zenkert, C. Weber, A. Klahold, M. Fathi, and K. Hahn, "Knowledge-based Production Documentation Analysis: An Integrated Text Mining Architecture," p. 4, 2018.
8. P. Subasic, H. Yin, and X. Lin, "Building Knowledge Base through Deep Learning Relation Extraction and Wikidata," p. 8, 2019.
9. L. Dietz, A. Kotov, and E. Meij, "Utilizing Knowledge Graphs for Text-Centric Information Retrieval," in *The 41st International ACM SIGIR Conference on Research & Development in Information Retrieval - SIGIR '18*, Ann Arbor, MI, USA, 2018, pp. 1387–1390, doi: 10.1145/3209978.3210187.
10. A. Tiwari, K. Rajesh, and P. Srujana, "Semantically Enriched Knowledge Extraction With Data Mining," *Int. J. Comput. Appl. Technol. Res.*, vol. 4, no. 1, pp. 7–10, Jan. 2015, doi: 10.7753/IJCATR0401.1002.
11. H. Al-Natsheh, "Text Mining Approaches for Semantic Similarity Exploration and Metadata Enrichment of Scientific Digital Libraries," p. 176, 2019.